\documentclass[conference]{IEEEtran}
\usepackage{cite}
\usepackage{amsmath,amssymb,amsfonts,mathtools}
\usepackage{bbm, bm}
\usepackage{algorithm}
\usepackage{algorithmic}
\usepackage{graphicx}
\usepackage{textcomp}
\usepackage{xcolor}
\usepackage{subcaption}
\usepackage{siunitx}
\usepackage{hyperref}

\newcommand{\grad}{\nabla}

\newcommand{\bI}{\mathbf{I}}

\newcommand{\bzero}{\mathbf{0}}

\newcommand{\bx}{\mathbf{x}}

\newcommand{\bz}{\mathbf{z}}

\newcommand{\bepsilon}{{\boldsymbol{\epsilon}}}

\def\BibTeX{{\rm B\kern-.05em{S  i\kern-.025em b}\kern-.08em
    T\kern-.1667em\lower.7ex\hbox{E}\kern-.125emX}}
    
\definecolor{mehdi}{RGB}{0,0,250}

\definecolor{Samad}{RGB}{0,250,0}

\definecolor{Matti}{RGB}{250,0,0}

\begin{document}

\title{Probabilistic
Constellation Shaping With Denoising Diffusion 
Probabilistic  
Models: A Novel Approach 
}

\author{
    \IEEEauthorblockN{Mehdi Letafati, Samad Ali, and Matti Latva-aho\\}
    \IEEEauthorblockA{\centering
    \begin{tabular}{c}
    {Centre for Wireless Communications,}
    {University of Oulu,}
      Oulu, Finland\\
    \texttt{\{mehdi.letafati, samad.ali, matti.latva-aho\}@oulu.fi}
	\end{tabular}
  \vspace{-1.0\baselineskip}
  }
\vspace{-1.5\baselineskip}
}

\maketitle

\begin{abstract} 
With the incredible results achieved from  
generative pre-trained transformers (GPT) and diffusion models,   
generative  AI (GenAI)  is envisioned to yield  remarkable breakthroughs in various industrial  and  academic domains.  
In this paper, 
we utilize 
denoising diffusion probabilistic models (DDPM), as one of the state-of-the-art generative models, 
for probabilistic constellation shaping in wireless communications.     
While the geometry of constellations is predetermined by the networking standards,   
probabilistic constellation shaping can help enhance the information rate and communication performance by designing  the probability of occurrence (generation) of constellation  symbols.  
{Unlike conventional methods that deal
with an optimization problem over the discrete distribution
of constellations, we take a radically different approach.  
Exploiting the “denoise-and-generate” characteristic of DDPMs,  the key idea is to learn how to generate constellation symbols out of noise, “mimicking’’ the way the receiver performs symbol reconstruction. By doing so, we make the 
constellation symbols sent by the transmitter, and what is inferred (reconstructed) at the receiver become as similar as possible.} 
Our  simulations  
show  that the proposed scheme  outperforms 
deep neural network (DNN)-based 
benchmark 
and uniform shaping, while providing \emph{network resilience} as well as \emph{robust out-of-distribution
performance} under low-SNR regimes and  non-Gaussian noise. Notably,   a threefold improvement in terms of mutual information  is achieved compared to  DNN-based approach  for 64-QAM geometry. 
\end{abstract}

\begin{IEEEkeywords}
AI-native wireless, diffusion models, generative AI,  network resilience, wireless AI. 
\end{IEEEkeywords}

\vspace{-1.0mm}
\section{Introduction}\label{sec:Intro}
\vspace{-1.5mm}
The emergence of  generative models has made a paradigm shift in the realm of artificial intelligence (AI)
towards generative AI (GenAI)-based systems \cite{Petar}.    
Innovative approaches in GenAI have attracted  significant  attention from both academia and industry, garnering    extensive   research and development efforts. In this regard, the evolution of diffusion models \cite{DM_Ho}, as the state-of-the-art family of generative models, is considered as one of the key factors in  the recent breakthroughs of GenAI. It has  showcased  remarkable results  
with famous solutions such as 
ImageGen
by  Google Brain 
and  DALL.E 2
by OpenAI,  to name a few.   
Through the lens of  data communication and networking, ``connected intelligence'' is envisioned  as the most significant driving force in the sixth generation (6G) of wireless communications \cite{3gpp, hexaX, Nokia_white, twelve_6G}. It is envisioned that machine learning (ML) and AI algorithms are  widely incorporated into wireless systems to realize  “AI-native”  systems.  This  highlights  the need for novel AI  solutions to be tailored for the emerging 6G scenarios.

\vspace{-1.0mm}
\subsection{Literature Review}
Although diffusion models have shown remarkable results  in  various  applications  within the computer science  community,  such as natural language processing (NLP), computer vision, and medical imaging \cite{DM_MRI}, 
 there are \emph{only a few papers in communication literature that have started looking into the applications  of  diffusion   models for wireless systems} \cite{DGM_Mag, CDDM, DM_for_E2EComm, CGM_ChanEst}.    
Notably, the incorporation  of diffusion models  into wireless communication problems is still in its infancy, and we hope that our paper sheds light on some of the possible directions.   

The authors in \cite{DGM_Mag} propose a workflow   for wireless network management via utilizing diffusion models,     
highlighting their exploration capability  for wireless network management.   
A preprint \cite{CDDM} employs  diffusion models to improve the  performance of receiver in terms of noise and channel estimation error removal.  
The authors employ an autoencoder (AE) in addition to their diffusion model.   However, 
the output signals of the encoder does not necessarily follow the standard shape of constellation symbols, making the scheme inapplicable to real-world wireless systems.    
Moreover, implementing  two different ML models, each with a distinct objective function 
can impose computational overhead  to the network.  
Denoising diffusion probabilistic model (DDPM) is utilized in \cite{DM_for_E2EComm} to  generate synthetic channel realizations for an AE-based end-to-end wireless system.  
The authors highlight the promising  performance  of diffusion models as an  alternative to generative adversarial network (GAN)-based models. They show that GANs  have  unstable training and less diversity in generation performance, because of  their ``adversarial'' training nature, while DDPMs maintain a more stable training process  and a better generalization during inference.  
In \cite{CGM_ChanEst}, noise-conditioned score networks 
are employed for  channel estimation in  multi-input-multi-output (MIMO) wireless communications. RefineNet neural architecture   
is implemented  to estimate the gradient of the log-prior of wireless channels.   
The results  imply  a competitive performance for in- and out-of-distribution (OOD) scenarios 
compared to GANs.

\vspace{-1mm}
\subsection{Our Work}  
With the aid of GenAI, our general goal is to take a step towards  an \emph{AI-native} system \cite{3gpp, hexaX,  Nokia_white}, in which  we can continuously design radio signals, adapt to changes, 
and  realize ``mutual  understanding'' between  communication parties, instead of blindly  
transmitting information symbols.   
In this paper,  we study the application of diffusion models,  the  state-of-the-art generative  model in GenAI literature, for probabilistic  constellation shaping in wireless communications.  \emph{To the best of our knowledge, this is the first paper that proposes diffusion models for   
constellation shaping in wireless communications.}

\vspace{-1.5mm}
{\textbf{\emph{Setting the Stage:}}}
The choice of constellations can significantly affect  the performance of communication systems.    
Recently, DNNs are proposed  for geometric shaping \cite{constell,waveform_learn,geom_arXiv}. They typically employ AEs and let the neural model learn constellation symbols for transmission.  This results in  arbitrary forms of constellation points that might not be compliant with wireless standards such as the 3rd Generation Partnership Project (3GPP) \cite{3gpp_constell}.        
In such scenarios, probabilistic constellation shaping can help enhance the information rate and decoding performance of communication systems \cite{constell}. It designs  the probability of occurrence (generation) of  constellation  symbols within the corresponding geometry.

\vspace{-1.5mm}
{\textbf{\emph{Our Contributions:}}}
Unlike previous works that try to
deal with the optimization of constellations over  discrete distributions via iterative optimization methods or deep neural networks (DNN) \cite{constell}, we offer  a radically different approach---we exploit  
the ``denoise-and-generate'' characteristic  of DDPMs for probabilistic shaping.    
First,  a DDPM  
is trained with the aim of learning the diffusion process for generating constellation symbols out of noise.  
Within each transmission slot (TS),  the transmitter 
runs the  model to probabilistically shape (generate) the constellation symbols  according to the signal-to-noise ratio (SNR) level.  
Intuitively, the goal is to do shaping in a way that the information-bearing constellation symbols generated at the transmitter, and what is inferred (reconstructed) at the receiver become as similar as possible,  resulting in as few mismatches between the communication parties as possible. 
To fulfill this requirement,  
the transmitter exploits the ``denoise-and-generate'' characteristic of DDPMs, and  ``mimics'' the way the receiver performs symbol reconstruction. 
(More details are provided in Section \ref{sec:SysMod}.)  
We show 
that our proposed approach  outperforms DNN-based scheme with trainable constellation layer and neural demapper  \cite{constell}. Notably, we show 
a threefold improvement in terms of mutual information metric compared to  DNN-based solution for $64$-QAM geometry.   
Our results also highlight that the proposed DDPM-based  scheme is  \emph{resilient} against  low-SNR regimes. 
We also demonstrate  a robust  OOD performance under non-Gaussian noise, compared to other benchmarks.   
 
\vspace{-0.5mm}
In what follows,  we first  introduce  DDPM framework in Section \ref{sec:DDPM}.   
System model and the proposed scheme are  introduced in Section \ref{sec:SysMod}.  Furthermore,  the neural architecture and the proposed  algorithms  for probabilistic constellation shaping  are addressed in this section.  
Numerical   results are studied in Section \ref{sec:Eval}, and  Section \ref{sec:concl} concludes the paper.\footnote{ 
Vectors and matrices are represented, respectively,  by bold lower-case and upper-case symbols. $|\cdot|$ and $||\cdot ||$ respectively denote the  absolute value of a scalar variable and the $\ell_2$ norm of a vector. Notation $\mathcal{N}(\mathbf{x}; \boldsymbol{\mu}, \mathbf{\Sigma})$  stands for the multivariate normal distribution  with mean vector $\boldsymbol{\mu}$ and covariance matrix $\mathbf{\Sigma}$ for a random vector $\mathbf{x}$. Similarly, complex normal distribution with the corresponding mean vector  and covariance matrix is denoted by $\mathcal{CN}(\boldsymbol{\mu}, \mathbf{\Sigma})$. Moreover, the expected value of a random variable (RV) is denoted by $\mathbb{E}\left[\cdot\right]$   Sets are denoted by calligraphic symbols.  $\bm 0$ and $\bf I$ respectively show all-zero vector and identity matrix of the corresponding size. Moreover, $[N]$,  (with $N$ as integer) denotes the set of all integer values from $1$ to $N$, and $\mathsf{Unif}[N]$ (for $N > 1$) denotes discrete uniform distribution  with samples between $1$ to $N$. Also,  $\delta(\cdot)$ denotes the Dirac function.
}

\vspace{-1mm}
\section{Preliminaries on DDPMs}\label{sec:DDPM}
\vspace{-1.5mm}
Diffusion models are a new class of state-of-the-art probabilistic  generative models inspired by non-equilibrium thermodynamics \cite{DM_Ho}. 
 Let ${\bf x}_0$  be a data sample from some  distribution ${ q}({\bf x}_0)$. For a finite number $T$ of time-steps, the forward diffusion process 
 $ q({\bf x}_t|{\bf x}_{t-1})$ 
 is defined by adding  Gaussian noise  
according to a ``variance schedule'' $0 < \beta_1 <  \beta_2 < \cdots <  \beta_T < 1$ at each time-step 
$t \in [T]$. 
 This is, 
\begin{align}
    q(\mathbf{x}_t \vert \mathbf{x}_{t-1}) 
    & \sim \mathcal{N}(\mathbf{x}_t; \sqrt{1 - \beta_t} \mathbf{x}_{t-1}, \beta_t\mathbf{I}),  \label{eq:fwd_diffusion}  
    \\
q(\mathbf{x}_{1:T} \vert \mathbf{x}_0) 
& = \prod^T_{t=1} q(\mathbf{x}_t \vert \mathbf{x}_{t-1}). \label{eq:diffusion_eqn}
\end{align} 
Invoking \eqref{eq:diffusion_eqn}, the data sample  gradually loses its distinguishable features as the time-step goes on, where with $T\!\rightarrow \!\infty$, ${\bf x}_T$ approaches an isotropic Gaussian distribution with covariance matrix ${\bf \Sigma}\!=\!\sigma^2\mathbf{I}$ for some $\sigma\!>\!0$ \cite{DM_Ho}. According to \eqref{eq:fwd_diffusion},  each new sample at time-step  $t$ can be drawn from a conditional Gaussian distribution with  mean vector  ${\mathbf \mu}_t = \sqrt{1 - \beta_t} \mathbf{x}_{t-1}$ and covariance matrix ${\bf \Sigma}^2_t = \beta_t \bf I$. Hence, the forward process is realized  by sampling from a Gaussian noise  $\bm{\epsilon}_{t-1} \sim {\cal N}(\bf 0, I)$  and setting 
\begin{align}\label{eq:fwd_sample_gen_diffusion}
	{\bf x}_t = \sqrt{1-\beta_t}&{\bf x}_{t-1} +\sqrt{\beta_t} {\bm{\epsilon}}_{t-1}.
\end{align} 
A useful property for  the forward  process in \eqref{eq:fwd_sample_gen_diffusion} is that we can sample ${\bf{x}}_t$ 
at any arbitrary time step $t$,  via recursively applying the reparameterization trick from  ML literature \cite{reparam_ML}. 
This results in the following formulation.  
\begin{align} 
\mathbf{x}_t  &= \sqrt{\bar{\alpha}_t}\mathbf{x}_0 + \sqrt{1 - \bar{\alpha}_t}\bm{\epsilon}_0,  \label{eq:xt_vs_x0} \\ 
    q({\bf x}_t|{\bf x}_0)&\sim\mathcal{N}\left({\bf x}_t;\sqrt{\bar{\alpha}_t}{\bf x}_0,(1-\bar{\alpha}_t)\mathbf{I}\right),\label{eq:xt_vs_x0_dist}
\end{align}
where $\bar{\alpha}_t\!=\!\prod_{i=1}^t(1-\alpha_i)$ and $\alpha_t=1-\beta_t$. 

Now the problem is to  reverse the process in \eqref{eq:xt_vs_x0} and sample from 
$q(\mathbf{x}_{t-1} \vert \mathbf{x}_t)$, so that we regenerate  the true samples from  some Gaussian noise  
$\mathbf{x}_T$.  
According to \cite{DM_Ho}, for $\beta_t$ 
small enough,  $q(\mathbf{x}_{t-1} \vert \mathbf{x}_t), \forall t \in [T]$  also follows Gaussian distribution. However,  we cannot easily estimate the distribution, since  it requires knowing the distribution of all possible data samples.    
Hence,  to approximate the conditional probabilities and  run the reverse diffusion process, we need to learn a probabilistic  model $p_{\bm \theta}(\mathbf{x}_{t-1} \vert \mathbf{x}_t)$, parameterized by ${\bm \theta}$.   
Accordingly, we can write    
\begin{align} 
p_{\bm \theta}(\mathbf{x}_{t-1} \vert \mathbf{x}_t) &\sim \mathcal{N}(\mathbf{x}_{t-1}; \boldsymbol{\mu}_{\bm \theta}(\mathbf{x}_t, t), \mathbf{\Sigma}_{\bm \theta}(\mathbf{x}_t, t)),  \label{eq:rev_proc_dist_conditional} 
\\ 
 p_{\bm \theta}(\mathbf{x}_{0:T}) &= p(\mathbf{x}_T) \prod^T_{t=1} p_{\bm \theta}(\mathbf{x}_{t-1} \vert \mathbf{x}_t).  \label{eq:rev_proc_dist_all} 
\end{align}
Now the problem simplifies  to learning the mean vector  $\boldsymbol{\mu}_{\bm \theta}(x_t,t)$ and the covariance matrix  $\mathbf{\Sigma}_{\bm \theta}(x_t,t)$ 
for the  probabilistic model $p_{\bm \theta}(\cdot)$, where a neural network (NN), with parameter ${\bm \theta}$,  can be trained to approximate (learn) the reverse process.  
We note that if we condition the reverse process  on ${\bf x}_0$, this conditional probability becomes  tractable \cite{DM_Ho}. 
Hence, when we have  ${\bf x}_0$ as a reference, we can take a small step backwards to generate data samples, and  the reverse step is formulated as $q({\bf x}_{t-1}|{\bf x}_t,{\bf x}_0)$.  
Mathematically, we have 
\begin{align}
    q(\mathbf{x}_{t-1} \vert \mathbf{x}_t, \mathbf{x}_0) &\sim \mathcal{N}(\mathbf{x}_{t-1}; \hspace{1.5mm} {\tilde{\boldsymbol{\mu}}}(\mathbf{x}_t, \mathbf{x}_0, t), {\tilde{\beta}_t} \mathbf{I}),   \label{eq:rev_conditioned_on_x0}
\end{align}
which is obtained by  utilizing  Bayes rule, where 
\begin{align}
    {\tilde{\boldsymbol{\mu}}}(\mathbf{x}_t, \mathbf{x}_0, t)
    &=\frac{\sqrt{\alpha_t}(1-\bar{\alpha}_{t-1})}{1-\bar{\alpha}_t}{\bf x}_t + \frac{\sqrt{\bar{\alpha}_{t-1}}
    \beta_t 
    }{1-\bar{\alpha}_t}{\bf x}_0, 
    \label{eq:mu_tilde}  \\ 
   {\tilde{\beta}_t} &=\frac{
   1-\bar{\alpha}_{t-1}}{1-\bar{\alpha}_t} \beta_t.\label{eq:beta_tilde}
\end{align}
Invoking \eqref{eq:beta_tilde}, one can infer that the covariance matrix in \eqref{eq:rev_conditioned_on_x0} has no learnable parameter. 
Hence, we simply need to learn the  mean vector ${\tilde{\boldsymbol{\mu}}}(\mathbf{x}_t, \mathbf{x}_0, t)$. 
To further simplify \eqref{eq:mu_tilde}, we note that  thanks to the reparameterization trick and with a similar approach to  \eqref{eq:xt_vs_x0},  we can express  ${\bf x}_0$ as follows.
\begin{align}
    \mathbf{x}_0 = \frac{1}{\sqrt{\bar{\alpha}_t}}(\mathbf{x}_t - \sqrt{1 - \bar{\alpha}_t}\bm{\epsilon}_t). \label{eq:x0_vs_xt}
\end{align}
Substituting ${\bf x}_0$ in \eqref{eq:mu_tilde} by \eqref{eq:x0_vs_xt} results in 
\begin{align}
    \begin{aligned}
\tilde{\boldsymbol{\mu}}(\mathbf{x}_t, \mathbf{x}_0, t) = 
{\frac{1}{\sqrt{\alpha_t}} \Big( \mathbf{x}_t - \frac{1 - \alpha_t}{\sqrt{1 - \bar{\alpha}_t}} \bm{\epsilon}_t \Big)}.
\end{aligned}
\end{align}
Now we can learn the conditioned probability distribution $p_{\bm \theta}(\mathbf{x}_{t-1} \vert \mathbf{x}_t)$
by training a NN  that approximates $\tilde{\boldsymbol{\mu}}(\mathbf{x}_t, \mathbf{x}_0, t)$.    
Therefore,  we simply need to set the approximated mean vector  $\boldsymbol{\mu}_{\bm \theta}(\mathbf{x}_t, t)$ to have the same form as the target mean vector $\tilde{\boldsymbol{\mu}}(\mathbf{x}_t, \mathbf{x}_0, t)$.  
Since $\mathbf{x}_t$  is available at time-step $t$, we can reparameterize the NN to make it approximate $\bm{\epsilon}_t$ 
 from the input $\mathbf{x}_t$. Compiling these facts  results in  
 \begin{align}\label{eq:mu_theta_to_learn}
\boldsymbol{\mu}_{\bm \theta}(\mathbf{x}_t, t) &=  {\frac{1}{\sqrt{\alpha_t}} \Big( \mathbf{x}_t - \frac{1 - \alpha_t}{\sqrt{1 - \bar{\alpha}_t}} \bm{\epsilon}_{\bm \theta}(\mathbf{x}_t, t) \Big)}, 
\end{align}
 where  $\bm{\epsilon}_{\bm \theta}(\mathbf{x}_t, t)$ denotes our NN.  
We can now define the loss function $\mathcal{L}_t$ for time-step $t \in [T]$, aiming  to minimize the difference between $\boldsymbol{\mu}_{\bm \theta}(\mathbf{x}_t, t)$ and  $\tilde{\boldsymbol{\mu}}(\mathbf{x}_t, \mathbf{x}_0, t)$.    
\begin{align}
\hspace{-2mm}
\mathcal{L}_t &=  
{\mathbb{E}}_{\begin{subarray}{l}t\sim {\mathsf{Unif}}[T]\\ {\bf x}_0\sim q({\bf x}_0) \\ \bm{\epsilon}_0\sim \mathcal{N}(\mathbf{0},\bf{I})\\ \end{subarray}}
\Big[\|\bm{\epsilon}_t - \bm{\epsilon}_{\bm \theta}(\mathbf{x}_t, t)\|^2 \Big]  \nonumber \\  
&=  {\mathbb{E}}_{\begin{subarray}{l}t\sim {\mathsf{Unif}}[T]\\ {\bf x}_0\sim q({\bf x}_0) \\ \bm{\epsilon}_0\sim \mathcal{N}(0,\bf{I})\\ \end{subarray}}  \Big[\|\bm{\epsilon}_t - \bm{\epsilon}_{\bm \theta}(\sqrt{\bar{\alpha}_t}\mathbf{x}_0 + \sqrt{1 - \bar{\alpha}_t}\bm{\epsilon}_t, t)\|^2 \Big]. \label{eq:loss_func}
\end{align}
Invoking \eqref{eq:loss_func}, 
at each time-step $t$,  the  DDPM model 
takes $\mathbf{x}_t$ as input  and returns the distortion components $\bm{\epsilon}_{\bm \theta}(\mathbf{x}_t,t)$.  Also,  $\bm{\epsilon}_t$ denotes the  diffused noise term   at time step $t$,

\begin{figure} 
\centering
\includegraphics
[width=2.45in, height=1.4in,trim={0.2in 0.08in 0.3in  0.09in},clip]{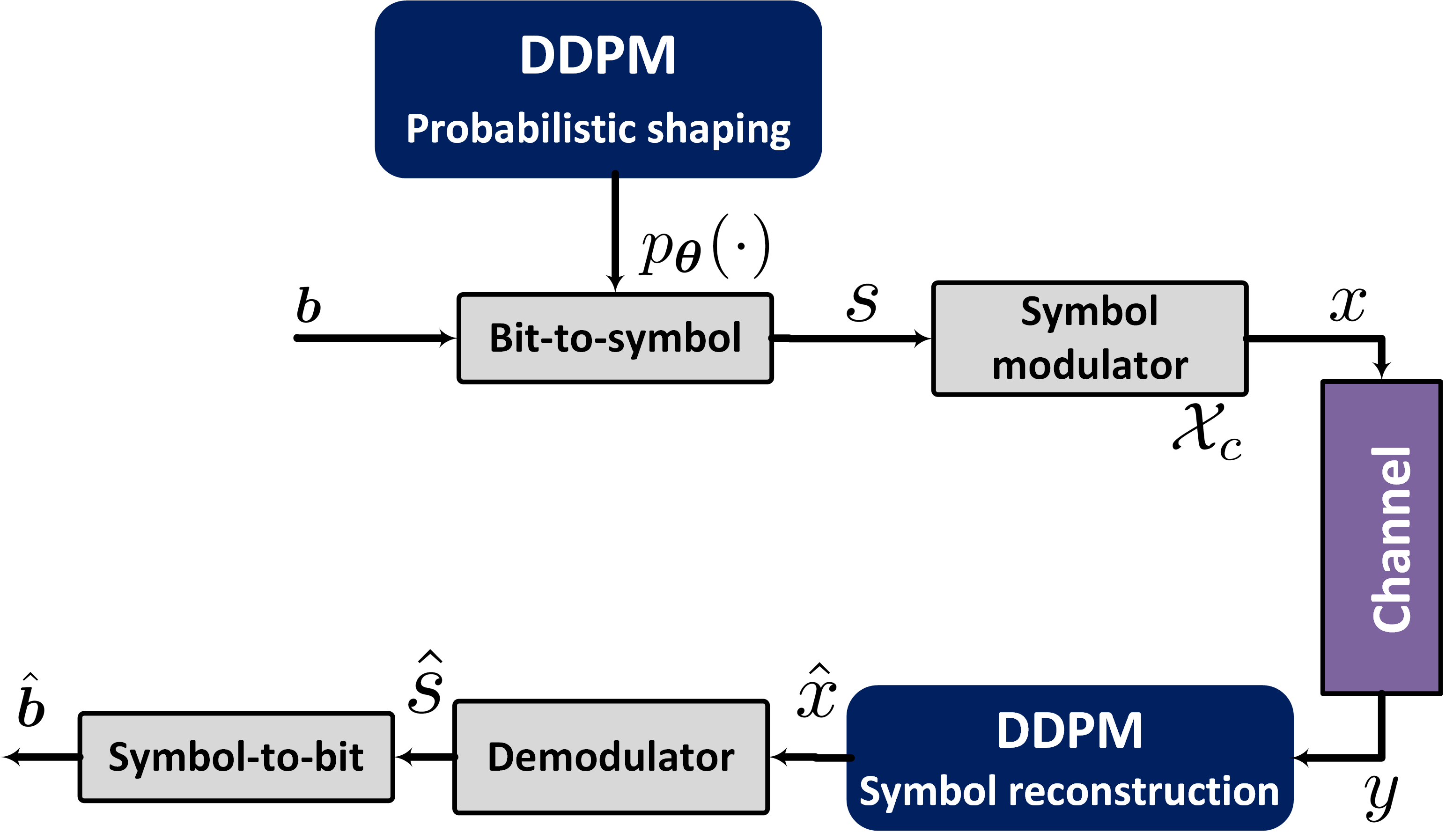}
\vspace{0mm}
\caption{\small System model overview. 
}
	\label{fig:SysMod}
 \vspace{-5mm}
\end{figure}

\section{System Model and Proposed Scheme}\label{sec:SysMod}
Fig. \ref{fig:SysMod} demonstrates the communication system under consideration.  
The system takes the information bitstream and maps it onto hypersymbols $s \in \mathcal{S}$, $\mathcal{S}  = \{1,\dots,M\}$, according to the learnable  distribution $p_{\bm \theta}(s)$  (parameterized by $\bm \theta$), where  $M$ denotes the modulation order.  
In this paper, $\bm \theta$ is realized by a DDPM, which is trained and shared  with the transmitter and receiver. 
The sequence of hypersymbols is then fed into a symbol modulator which maps each symbol $s$ into a constellation point $x \in \mathcal{X}_c$, with ${\cal X}_c$ showing the set of constellation points. 
Each symbol is  generated  according to the distribution $p_{\bm \theta}(s)$.    
In other words, the frequency of sending a bitstream over the constellation point $x = g(s)$  corresponds to the parametric distribution $p_{\bm \theta}(s)$, where $g$ denotes the  modulation functionality.  Accordingly,    
we have 
\begin{align}\label{eq:pX}
p_{\bm \theta}(x) = \sum_{s \in \mathcal{S}} \delta\left(x - g(s)\right)p_{\bm \theta}(s), \quad \forall x \in \mathcal{X}_c. 
\end{align}
As motivated in Section \ref{sec:Intro}, our focus in this work is on probabilistic shaping, and hence, constellation geometry is determined in advance according to network  configurations. 
Thus, the design of modulator function $g(\cdot)$ is not  of interest in this work, and we adhere to  standard constellation schemes, such as QAM, in order to propose a system which is compliant with the real-world communication systems. 
In addition, similar to \cite{constell}, we also assume that the bits-to-symbols mapper is  known. 
Hence, we have a one-to-one mapping between the constellation point $x$
and the information  symbol $s$, and the transmitter's output  is directly   sampled from  $p_{\bm \theta}(\cdot)$.   
Information-bearing signal $x$ is then sent over the communication  channel, and the channel output $y$ is observed at the receiver. 
Then the receiver needs to reconstruct the transmitted symbols by approximating the posterior distribution  $p(s|y)$ given the channel output.   
To do so,  the receiver leverages the trained DDPM  
and  maps each received sample $y$ to a probability distribution  over the set of symbols $\cal S$.    
Having this approximation,  symbols can be obtained  at the receiver's  de-modulator, and  
the information bits 
can be reconstructed  
using the prevalent symbol-to-bit  mappers.

\vspace{0mm}
\subsection{Proposed Approach}\label{subsec:proposed_approach} 
\subsubsection*{\textbf{Setting the Stage}}
Intuitively speaking,  the goal is to probabilistically  shape the constellation symbols by finding a proper $p_{\bm \theta}(\cdot)$, such that 
the information-bearing symbols 
sent by 
the transmitter,  
and what is inferred 
at the receiver 
become as similar as possible,\footnote{This ``similarity'' will be quantitatively evaluated in Section \ref{sec:Eval} using the widely-adopted   mutual information metric.}
resulting in as few mismatches between the communication parties as possible. 
This fact, together with the characteristic of  diffusion models to ``denoise-and-generate'',  motivates us to propose the following DDPM-based approach for probabilistic constellation shaping.   
The key idea to fulfill the desired  similarity  is that 
the transmitter ``mimics'' the way the receiver would perform  the reconstruction of symbols. 
Hence, the transmitter probabilistically generates  the 
constellations in a way that would be similar to  
the process of denoising and reconstruction (regeneration) at the receiver.       
This also helps facilitate having 
``mutual understanding'' of how to map and de-map the information symbols over time, realizing kind of \emph{native intelligence} among communication parties. Motivated by these facts, our step-by-step solution  can be elaborated on as follows.

\subsubsection{DDPM Training}\label{step1} 
A  DDPM  is trained based on the loss function given in \eqref{eq:loss_func}. This  corresponds to training the parameter $\bm \theta$ for our probabilistic shaping scheme  in \eqref{eq:pX}.   
The goal is to train a diffusion process to generate  constellation symbols  (with the pre-determined geometry) out of noise. The  process
is summarized in Algorithm \ref{alg:trainAlg}, which is inspired by the seminal paper of DDPM by Ho \emph{et. al}, 2020 \cite{DM_Ho}.  
{Training can be carried out in a central cloud, or an edge server  
and then downloaded by the communication entities.}
{The trained model is  deployed at the  transmitter and the receiver.} 

\subsubsection{Link quality estimation using channel SNR}
Within each TS,  the transmitter  first  estimates the quality of communication link. This can be carried out  using  the pilot signals sent by the destination node,  at the beginning of each TS, and the SNR level of communication channel  can be calculated 
\cite{Derrick_MIMO}.

\subsubsection{Probabilistic shaping}
The trained DDPM is run at the transmitter to probabilistically shape (generate) the constellation symbols according to the channel SNR.    
To do so, the transmitter first 
takes $N_s$ samples
from the set of constellation symbols $\mathcal{X}_c$ uniformly at random.\footnote{The sample size $N_s$ can be regarded as the  number of observations 
to form (generate) the empirical distribution of our probabilistic shaping.}  
The goal is not to 
uniformly  
map information symbols to constellation points.   
Rather, we aim to generate the constellation symbols in a way that the information-bearing constellation symbols sent by the transmitter, and what is inferred (reconstructed) at the receiver become as similar as possible.  
For instance, when the communication channel is experiencing high levels of noise, i.e.,  in low-SNR regime,  we intuitively expect that  most often, the  receiver would be able to  decode the symbols corresponding to the 
points that are relatively far from each other in the constellation geometry, while the other points are prone to being decoded incorrectly. 
Thus, the transmitter wishes to probabilistically  reshape the constellation symbols in a way that would be 
straightforward to  
denoise and reconstruct (regenerate) at the  receiver.  
To do so, the transmitter samples $N_s$ random  
noise with  
average power $\delta^2$, and injects them  to the uniformly-sampled symbols.  
The power of synthetic noise,   $\delta^2$,   is calculated according  to the channel SNR, $\Gamma$,  which is obtained at Step 2, and can be formulated  as 
$ \delta ^ 2 =  P {10^{{ \Gamma}/{10}}},$ 
where $P$ denotes the average transmit power. 
The  
noisy 
version of samples is then fed into the trained DDPM,  and   the  reverse diffusion process is run to denoise and  generate  symbols out of the synthetically-noisy samples.
In other words, the transmitter tries to mimic the way the receiver  performs the reconstruction  (regeneration)  of symbols, when it receives noisy symbols.
The distribution of the generated  samples at the output of the DDPM block is considered 
as the output  probabilistic  constellations onto which the information symbols are mapped to be sent.

The overall algorithm is proposed in Algorithm \ref{alg:sampling_Tx},    
where the main loop corresponds to the reverse diffusion process from time-step $T$ to $1$, according to \ref{eq:mu_theta_to_learn}.   
Also, $\mathtt{proj}_{\mathcal{S}}(\mathbf{x})$ stands for the projection operator, which  maps  the elements  of vector $\mathbf{x}$ onto the nearest elements in the set $\mathcal{S}$.  Moreover,  $\mathtt{count}(\mathbf{x}, \mathcal{S})$ outputs a vector with 
size $|\mathcal{S}|$,  
with elements representing  the number of occurrences of the elements of set $\cal S$ in vector $\mathbf{x}$.    
Notably,  $\boldsymbol{\psi}$ in Algorithm \ref{alg:sampling_Tx}  denotes the probabilistically-shaped constellation points at the output of the transmitter's DDPM block, and $p_{\bm \theta}$ stands for the corresponding distribution inferred by the diffusion model.        
 
 \begin{figure}[t]
\vspace{-3mm}
\begin{algorithm}[H]
\small
\hspace*{0.02in} {\bf {Hyper-parameters:}}
	{Number of time-steps $T$, neural architecture $\boldsymbol{\epsilon}_{\bm \theta}(\cdot, t)$, variance schedule   $\beta_t$, and $\bar{\alpha}_t, \forall t \in [T]$.} \\
    \hspace*{0.02in} {\bf {Input:}}
	{Training samples from   the constellation geometry $\mathcal{X}_c$.} \\
	\hspace*{0.02in} {\bf {Output:}} {Trained neural model for DDPM.}
	\caption{\small Training algorithm of DDPM}
	\label{alg:trainAlg}
	\begin{algorithmic}[1] 
 \small
    \WHILE {the stopping criteria are not met}
    \STATE Randomly sample $\mathbf{x}_0$ from  $\mathcal{X}_c$
    \STATE Randomly sample $t$ from $\mathsf{Unif}[T]$ 
    \STATE Randomly sample $\boldsymbol{\epsilon}$ from $\mathcal{N}(\mathbf{0},\mathbf{I})$ 
      \STATE Take gradient descent step on
      \STATE $\qquad {\grad}_{\boldsymbol{\theta}} \left\| \bepsilon - \bepsilon_{\bm \theta}(\sqrt{\bar\alpha_t} \bx_0 + \sqrt{1-\bar\alpha_t}\bepsilon, t) \right\|^2$
    \ENDWHILE
	\end{algorithmic}
\end{algorithm}  
\vspace{-8mm}
\end{figure}

 \begin{figure}[t]
\vspace{-3mm}
\begin{algorithm}[H]
\small
\hspace*{0.02in} {\bf {Hyper-parameters:}}
	{Number of time-steps $T$, trained neural model $\bm \theta$, constellation geometry $\mathcal{X}_c$} \\
    \hspace*{0.02in} {\bf {Input:}}
	{Channel SNR $\Gamma$.} 
  \caption{\small DDPM sampling: Probabilistic shaping at transmitter}  \label{alg:sampling_Tx}
  \small
  \begin{algorithmic}[1]
    \vspace{0.0in}
    \STATE  Randomly sample $\widetilde{\mathbf{x}}_s$, with size $N_s$, from set  $\mathcal{X}_c$ 
    \STATE Randomly sample $\widetilde{\mathbf{n}}$, with size $N_s$,  from $\mathcal{N}(\mathbf{0}, \mathbf{I})$ 
    \STATE $\widetilde{\mathbf{y}}_r = \widetilde{\mathbf{x}}_s + \delta  \widetilde{\mathbf{n}}$
    \STATE $\bx_T = \widetilde{\mathbf{y}}_{ r}$ 
    \FOR{$t=T, ... , 1$}
      \STATE $\bz \sim \mathcal{N}(\bzero, \bI)$ if $t > 1$, else $\bz = \bzero$
      \STATE $\bx_{t-1} = \frac{1}{\sqrt{\alpha_t}}\left( \bx_t - \frac{1-\alpha_t}{\sqrt{1-\bar\alpha_t}} \bepsilon_{\bm \theta}(\bx_t, t) \right) + \sqrt{1-\alpha_t} \bz$
    \ENDFOR 
    \STATE $\boldsymbol{\psi} = \mathtt{proj}_{\mathcal{X}_c}(\mathbf{x}_0)$
    \STATE $\boldsymbol{c} \hspace{1mm} =  \mathtt{count} \left(\boldsymbol{\psi}, \mathcal{X}_c\right)$  
    \STATE \textbf{return} $p_{\bm \theta} = 
     {\boldsymbol{c}}/{N_s}$
    \vspace{0.0in}
  \end{algorithmic}
\end{algorithm}
\vspace{-8mm}
\end{figure} 

\subsubsection{Symbol reconstruction at the receiver} 
After  generating constellation symbols, 
 information signals are sent according to the probabilistic constellations. The symbols are  received by the receiver.  Then 
the receiver runs the diffusion model 
to reconstruct (regenerate) the symbols from the received noisy signals.   The corresponding  algorithm for this step is proposed in Algorithm \ref{alg:sampling_Rx}. 
Starting from the received batch of noisy symbols, denoted by $\mathbf{y}_{r}$, for each time step $t\in\{T, T-1, \ldots, 1\}$,  the NN outputs $\bm{\epsilon_{\bm \theta}}(\hat{\mathbf{x}}_t, t)$ to approximate  the residual noise within the batch of symbols, and the sampling algorithm is run according to Line $4$ of the algorithm, in order  to sample $\hat{\mathbf{x}}_{t-1}$. The process is executed for $T$ steps.\footnote{We emphasize that within each TS, while the channel coherence time is respected,  the channel SNR remains unchanged 
compared to the one that is utilized by the transmitter for the  constellation shaping.}

\begin{figure}[t]
\vspace{-4mm}
\begin{algorithm}[H]
\small
\hspace*{0.02in} {\bf {Hyper-parameters:}}
	{Number of time-steps $T$, trained neural model $\bm \theta$, constellation geometry $\mathcal{X}_c$} \\
    \hspace*{0.02in} {\bf {Input:}}
	{Received signal ${\mathbf{y}}_r$} 
  \caption{\small   DDPM sampling: Symbol reconstruction at receiver}  \label{alg:sampling_Rx}
  \begin{algorithmic}[1]
    \vspace{0.0in}
    \STATE $\hat{\mathbf{x}}_T = \mathbf{y}_{r}$ 
    \FOR{$t=T, ... , 1$}
      \STATE $\bz \sim \mathcal{N}(\bzero, \bI)$ if $t > 1$, else $\bz = \bzero$
      \STATE $\hat{\mathbf{x}}_{t-1} = \frac{1}{\sqrt{\alpha_t}}\left( \hat{\mathbf{x}}_t - \frac{1-\alpha_t}{\sqrt{1-\bar\alpha_t}} \bepsilon_{\bm \theta}(\hat{\mathbf{x}}_t, t) \right) + \sqrt{1-\alpha_t} \bz$
    \ENDFOR
    \STATE \textbf{return} $\mathtt{proj}_{\mathcal{X}_c}(\hat{\mathbf{x}}_0)$  
  \end{algorithmic}
\end{algorithm}
\vspace{-8mm}
\end{figure}

\section{Evaluations}\label{sec:Eval}
In this section, we carry out  numerical evaluations, in order  to highlight the performance of the proposed scheme compared to other benchmarks.  Specifically, we show that our DDPM-based approach  achieves  a threefold improvement in terms of mutual information compared to  DNN-based solution for $64$-QAM geometry.    We also show that the proposed  DDPM  can provide  \emph{native  resilience} for the communication system under low-SNR regimes and  non-Gaussian noise.

We employ a NN  comprised of $3$  hidden linear layers 
each of which has $128$ neurons with softplus activation functions. 
The output layer is a simple  linear layer with the same shape as  input. Inspired by the Transformer paper \cite{attention}, 
we share the   parameters of the NN across time-steps via  multiplying the embeddings of time-step and incorporating them into the  model. 
For training the diffusion model, we use adaptive moment estimation (Adam) optimizer with  learning rate  $\lambda = 10^{-3}$. We consider QAM geometry as a   widely-adopted constellation format in wireless networks \cite{twelve_6G, hexaX, constell}.  Moreover, we set  $T=100$, and the stopping criterion in Algorithm \ref{alg:trainAlg} is met when reaching the maximum number of epochs \cite{CDDM, DM_for_E2EComm, CGM_ChanEst}, which is set to $1000$ epochs.  

\begin{figure}
\centering
    \begin{subfigure}{0.5\textwidth}
        \centering
        \includegraphics[width=\linewidth, trim={1.5in 0.0in 1.0in  0.1in},clip]{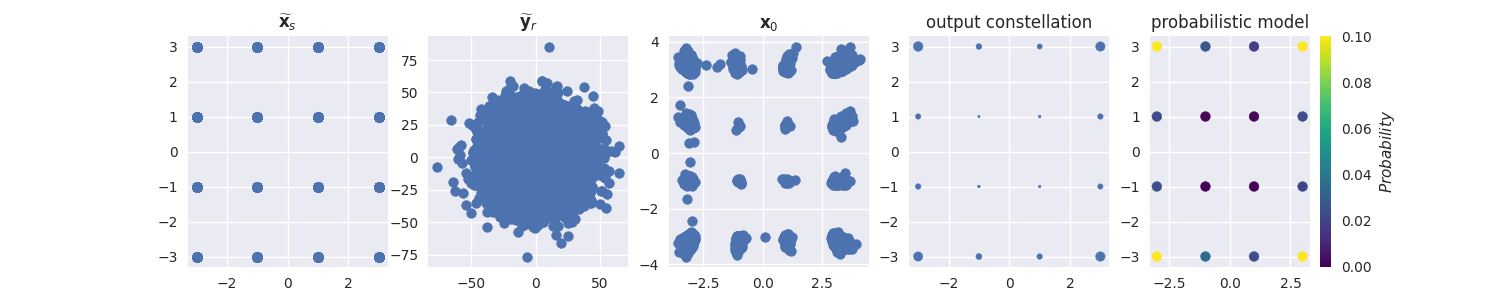} 
    \end{subfigure}
    \begin{subfigure}{0.5\textwidth}
        \centering
        \includegraphics[width=\linewidth, trim={1.5in 0.0in 1.0in  0.0in},clip]{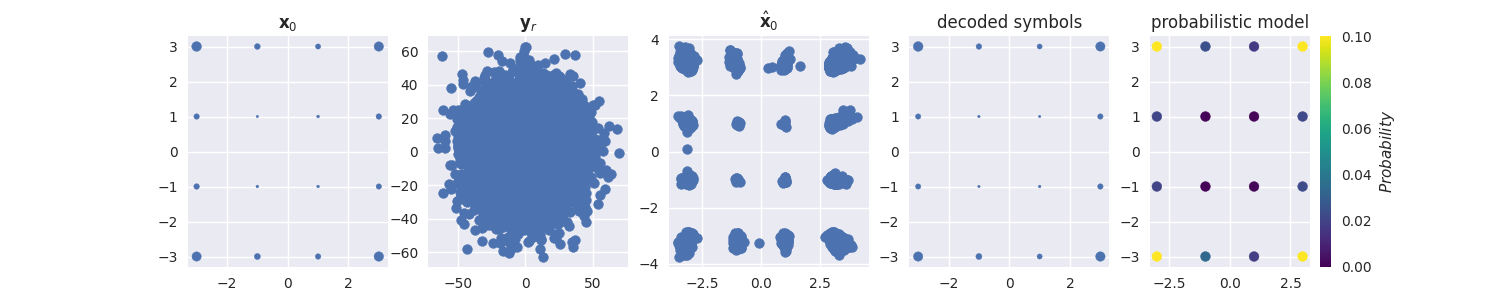} 
    \end{subfigure}
    \begin{subfigure}{0.5\textwidth}
        \centering
        \includegraphics[width=\linewidth, trim={1.5in 0.0in 1.0in  0.0in},clip]{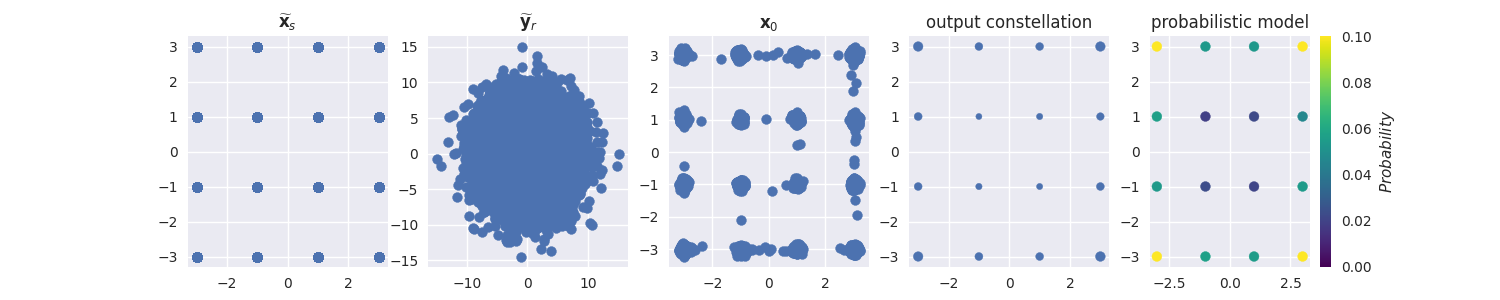} 
    \end{subfigure}
    \begin{subfigure}{0.5\textwidth}
        \centering
        \includegraphics[width=\linewidth, trim={1.5in 0.0in 1.0in  0.0in},clip]{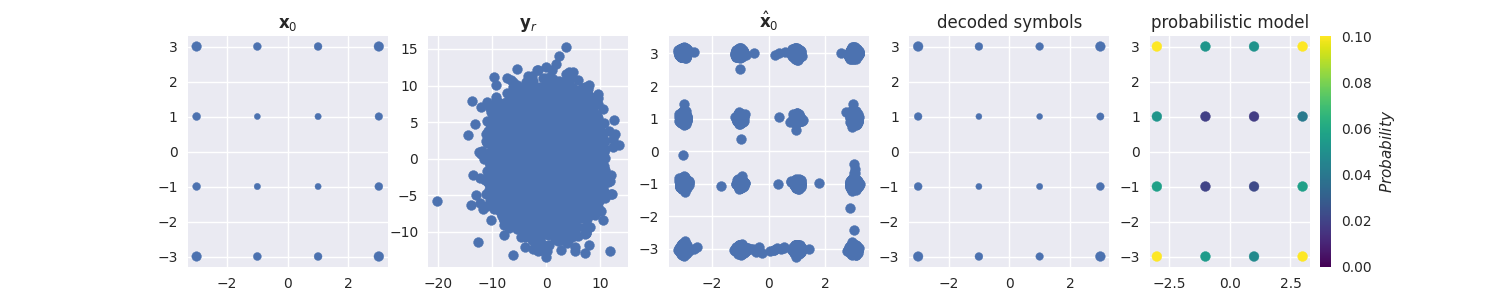} 
    \end{subfigure}
    \begin{subfigure}{0.5\textwidth}
        \centering
        \includegraphics[width=\linewidth, trim={1.5in 0.0in 1.0in  0.0in},clip]{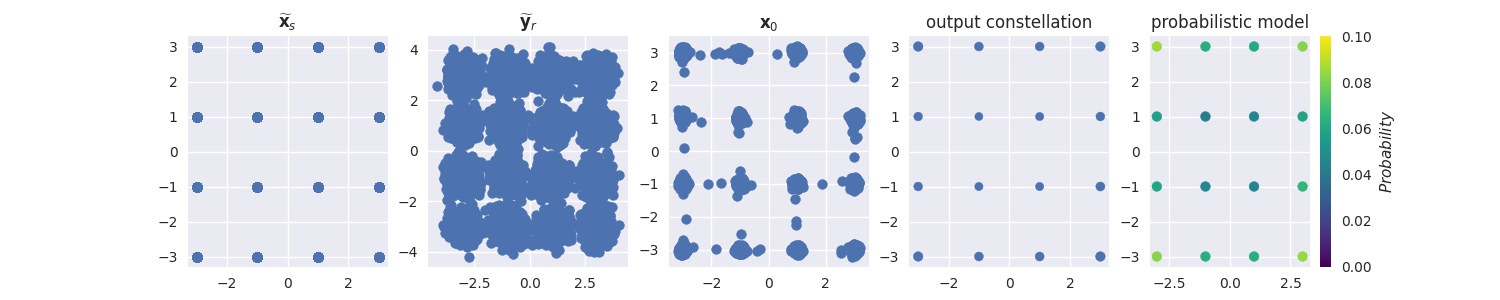} 
    \end{subfigure}
    \begin{subfigure}{0.5\textwidth}
        \centering
        \includegraphics[width=\linewidth, trim={1.5in 0.0in 1.0in  0.0in},clip]{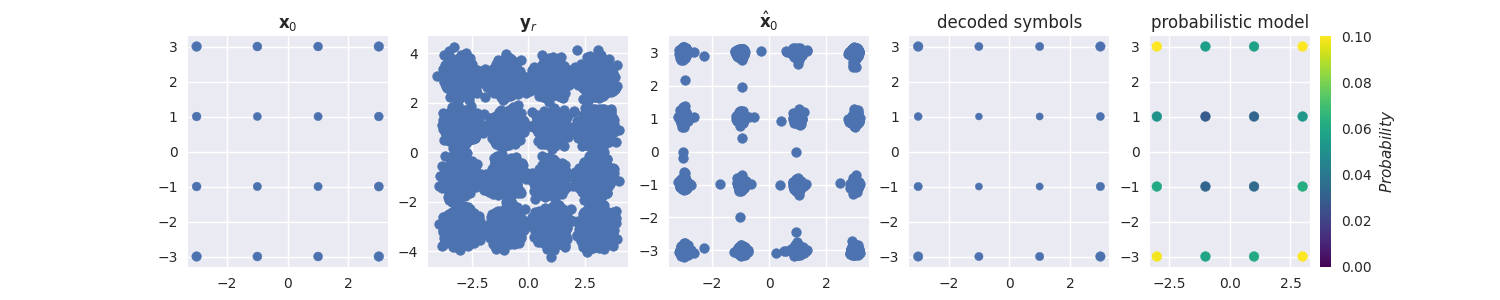} 
    \end{subfigure}
    \caption{\small Generation process at the transmitter, and the reconstruction at receiver  
    for SNR values set to  $-25$, $-10$, and $10$ dB, respectively.} 
    \label{fig:denoise}
    \vspace{0mm}
\end{figure}

Fig. \ref{fig:denoise}, demonstrates  data visualization for sampling phase, corresponding to Algorithms \ref{alg:sampling_Tx} and \ref{alg:sampling_Rx}. 
The first row   
corresponds to the constellation generation steps that are performed  at the transmitter (Algorithm \ref{alg:sampling_Tx}), and the second row corresponds to the reconstruction at the receiver (Algorithm \ref{alg:sampling_Rx}).   This is repeated for different SNRs to visualize the constellation shaping  performance of our DDPM under different levels of noise.       
Comparing the output of our probabilistic constellation generation algorithm (the first row) and the reconstructed symbols at the receiver (the second row), we can observe that the idea of synthetically  mimicking the functionality of receiver for shaping  the constellation symbols (addressed in Section \ref{subsec:proposed_approach}) has helped the transmitter generate symbols that are quite similar to the ones that are actually reconstructed by the receiver. This can improve the communication performance by   decreasing the mismatch between the way the transmitter decides to convey  the information, and the way the receiver  decodes the symbols. This ``similarity'' is quantitatively measured  in terms of mutual information  in the next figure. 
According to the figure,  when the communication system is experiencing low-SNR regimes,   the probabilistic model 
demonstrates a non-uniform distribution over constellation points, with higher probabilities assigned to the  points that are at the furthest distance from each other in the constellation geometry. This is aligned with what we intuitively expect  from a communication system under low-SNR regimes  to   frequently  map information bits to constellation symbols that are far apart from each other.    
Increasing the SNR,  
the probabilistic shaping tends to uniform distribution, which is also aligned with one's intuition about communication systems.

\begin{figure}[t] 
\centering
\includegraphics
[width=3.0in,height=2.0in,trim={0.4in 0.2in 0.4in  0.5in},clip]{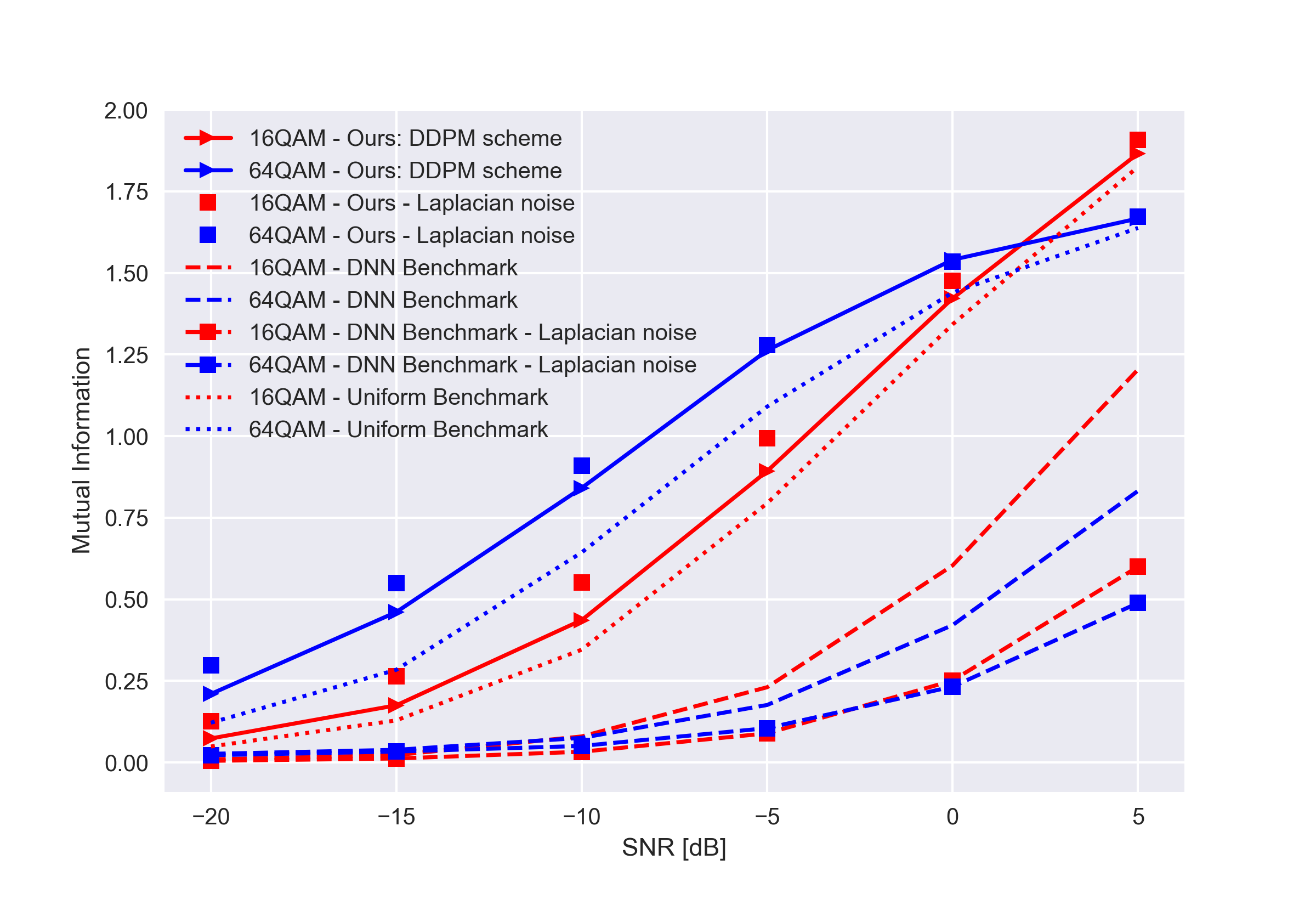}
\vspace{-2mm}
\caption{\small Mutual information  between the   
generated symbols at transmitter
and  the reconstructed ones at the receiver.   
}
	\label{fig:MI}
 \vspace{-5.5mm}
\end{figure}

Fig. \ref{fig:MI} demonstrates the mutual information  between the  probabilistically-generated symbols at the output of  transmitter  (i.e., the channel input),  and  the reconstructed ones at the receiver. 
The mutual information metric can be interpreted as the quantitative measure to study the ``mutual similarity'' among communication parties as discussed in Section \ref{subsec:proposed_approach}.  
For this experiment, we consider both cases of additive white Gaussian noise (AWGN) channel and also  non-Gaussian noise to show the OOD performance of our scheme.    
For the benchmark, we consider a DNN model with trainable constellation layer and neural demapper as proposed in  \cite{constell}.
The DNN benchmark has three linear layers with $64$ neurons at hidden layers and rectified linear unit (ReLU) activation functions, and we considered $5000$ training iterations  with Adam optimizer.    
Fig. \ref{fig:MI} clearly highlights the performance of our DDPM-based model in low-SNR regimes.  Notably,  although the  DNN benchmark  
does not show any noticeable  performance in SNR ranges below $-5$ dB,  even for the more straightforward  scenario of $16$-QAM (which is supposed to be less prone to errors and mismatches than 64-QAM case), our scheme achieves mutual information of around $1.25$ bits for 64-QAM geometry, and $1$ bit for 16-QAM geometry, respectively.  
Moreover, our scheme shows a threefold improvement  compared to  DNN-based benchmark  for 64-QAM geometry and $0$ dB SNR.  
These results clearly highlight that our main goal in realizing the ``mutual understanding'' among communication parties has  been successfully achieved, and thanks to this understanding,  the system is resilient under low-SNRs.     
To show the robustness of our scheme in  OOD performance, we study the scenario of communication  channels with non-Gaussian noise.     
We consider additive Laplacian noise with the same variance as that of AWGN scenario  as our benchmark \cite{twelve_6G}.   Remarkably, although we do not re-train our diffusion model with  Laplacian noise,  the performance of  our DDPM-based approach does not change (even becomes better) 
under this non-Gaussian assumption (which is not seen during training), and the resultant  mutual information curves follow the case of in-distribution scenario.  However, 
the DNN benchmark experiences performance degradation under non-Gaussian assumption,  although  we also re-trained it with Laplacian noise.  
In Fig. \ref{fig:MI}, we also study another benchmark, which is uniform shaping. 
We examine this benchmark by disabling the DDPM block at the transmitter.  We can see from the figure   that still,  the employed  DDPM at receiver can outperform conventional DNN benchmark. 
Notably, the figure implies that our probabilistic shaping model with DDPMs employed  at both the transmitter and the receiver outperforms the naive scheme of uniform shaping.  
This gap also increases when  considering higher modulation orders, as  it becomes more important to realize somewhat  mutual understanding and smartly shaping the higher order  constellation symbols.

\section{Conclusions}\label{sec:concl} 
In this paper,  we studied  the application of DDPMs  in  probabilistic constellation shaping for wireless communications.   
We exploited  the ``denoise-and-generate'' characteristic  of DDPMs.    
The transmitter runs the  model to probabilistically  shape (generate) the constellation symbols  and the receiver regenerates (reconstructs) the symbols from the received noisy signals.   
The key idea  was that   the transmitter mimics the way the receiver would do to reconstruct (regenerate) the symbols out of noisy signals, realizing   
``mutual understanding'' to reduce  the mismatch among communication parties.   
Our results 
highlighted the performance of our scheme  compared to  DNN-based demapper, while providing \emph{network resilience} under  low-SNR regimes and  non-Gaussian noise.

\def\baselinestretch{0.92}


\end{document}